\documentclass{PoS}

\title{The lunar Askaryan technique: a technical roadmap }
\ShortTitle{The lunar Askaryan technique: a technical roadmap }

\newcommand{\nijmegen}{Dept.\ of Astrophysics/IMAPP, Radboud Univ.\ Nijmegen, 6500 GL Nijmegen, The Netherlands}
\newcommand{\astron}{Netherlands Institute for Radio Astronomy (ASTRON), 7990 AA Dwingeloo, The Netherlands}
\newcommand{\karlsruhe}{IKP, Karlsruhe Institut f\"ur Technologie, Postfach 3640, 76021 Karlsruhe, Germany}
\newcommand{\erlangen}{ECAP, Univ.\ of Erlangen-Nuremberg, 91058 Erlangen, Germany}
\newcommand{\groningen}{KVI-CART, Univ.\ of Groningen, 9747 AA Groningen, The Netherlands}
\newcommand{\manchester}{School of Physics \& Astronomy, Univ.\ of Manchester, M13 9PL, United Kingdom}
\newcommand{\adelaide}{School of Chemistry \& Physics, Univ.\ of Adelaide, SA 5005, Australia}
\newcommand{\iowa}{Dept.\ of Physics \& Astronomy, Univ.\ of Iowa, IA 52242, USA}
\newcommand{\atnf}{CSIRO Astronomy \& Space Science, NSW 2122, Australia}
\newcommand{\lebedev}{LPI, Russian Academy of Sciences, Moscow Region 142290, Russia}
\newcommand{\santiago}{Depto.\ de F\'isica de Part\'iculas \& Instituto Galego de F\'isica de Altas Enerx\'ias, Univ.\ de Santiago de Compostela, 15782 Santiago de Compostela, Spain}
\newcommand{\brussels}{Astrophysical Institute, Vrije Univ.\ Brussel, Pleinlaan 2, 1050 Brussels, Belgium}
\newcommand{\iihe}{Interuniv.\ Institute for High-Energy, Vrije Univ.\ Brussel, Pleinlaan 2, 1050 Brussels, Belgium}

\author{
 J.D.~Bray$^1$, J.~Alvarez-Mu\~niz$^2$, S.~Buitink$^3$, R.D.~Dagkesamanskii$^4$, R.D.~Ekers$^5$, H.~Falcke$^{6,7}$, K.G.~Gayley$^8$, T.~Huege$^9$, C.W.~James$^{10}$, M.~Mevius$^{11}$, R.L.~Mutel$^8$, R.J.~Protheroe$^{12}$, O.~Scholten$^{11,13}$, R.E.~Spencer$^{1}$ and S.~ter~Veen$^3$\\
 \llap{$^1$}\manchester \\
 \llap{$^2$}\santiago \\
 \llap{$^3$}\brussels \\
 \llap{$^4$}\lebedev \\
 \llap{$^5$}\atnf \\
 \llap{$^6$}\nijmegen \\
 \llap{$^7$}\astron \\
 \llap{$^8$}\iowa \\
 \llap{$^9$}\karlsruhe \\
 \llap{$^{10}$}\erlangen \\
 \llap{$^{11}$}\groningen \\
 \llap{$^{12}$}\adelaide \\
 \llap{$^{13}$}\iihe \\
 E-mail: \email{justin.bray@manchester.ac.uk}
}

\abstract{
 The lunar Askaryan technique, which involves searching for Askaryan radio pulses from particle cascades in the outer layers of the Moon, is a method for using the lunar surface as an extremely large detector of ultra-high-energy particles.  The high time resolution required to detect these pulses, which have a duration of around a nanosecond, puts this technique in a regime quite different from other forms of radio astronomy, with a unique set of associated technical challenges which have been addressed in a series of experiments by various groups.  Implementing the methods and techniques developed by these groups for detecting lunar Askaryan pulses will be important for a future experiment with the Square Kilometre Array (SKA), which is expected to have sufficient sensitivity to allow the first positive detection using this technique.

 Key issues include correction for ionospheric dispersion, beamforming, efficient triggering, and the exclusion of spurious events from radio-frequency interference.  We review the progress in each of these areas, and consider the further progress expected for future application with the SKA.
}

\FullConference{The 34th International Cosmic Ray Conference \\
		 30 July- 6 August, 2015\\
		 The Hague, The Netherlands}

\begin{document}

\newcommand{\aap}{A\&A}                  
\newcommand{\aapr}{A\&A Rev.}            
\newcommand{\aaps}{A\&AS}                
\newcommand{\aipcs}{AIP Conf.\ Series}   
\newcommand{\aj}{AJ}                     
\newcommand{\ajph}{Australian J.\ Phys.} 
\newcommand{\alet}{Astro.\ Lett.}        
\newcommand{\anchem}{Analytical Chem.}   
\newcommand{\ao}{Applied Optics}         
\newcommand{\apj}{ApJ}                   
\newcommand{\apjl}{ApJ Lett.}                  
\newcommand{\apjs}{ApJS}                 
\newcommand{\app}{Astropart.\ Phys.}     
\newcommand{\apss}{Ap\&SS}               
\newcommand{\apssproc}{Ap\&SS\ Proc.}    
\newcommand{\araa}{ARA\&A}               
\newcommand{\arep}{Astron.\ Rep.}        
\newcommand{\arxiv}{ArXiv e-prints}      
\newcommand{\aspacer}{Adv.\ Space Res.}  
\newcommand{\aspconf}{Astron.\ Soc.\ Pac.\ Conf.} 
\newcommand{\atel}{ATel}                 
\newcommand{\azh}{AZh}                   
\newcommand{\baas}{BAAS}                 
\newcommand{\bell}{Bell Systems Tech.\ J.} 
\newcommand{\cpc}{Comput.\ Phys.\ Commun.} 
\newcommand{\crp}{Compt.\ Rend.\ Phys.}  
\newcommand{\cosres}{Cosm.\ Res.}        
\newcommand{\dans}{Dokl.\ Akad.\ Nauk SSSR} 
\newcommand{\easconf}{EAS Pub.\ Series}  
\newcommand{\elec}{Electronics}          
\newcommand{\epjwoc}{EPJ Web of Conf.}   
\newcommand{\epsl}{Earth and Plan.\ Sci.\ Lett.} 
\newcommand{\expa}{Exp.\ Astron.}        
\newcommand{\gca}{Geochim.\ Cosmochim.\ Acta} 
\newcommand{\grl}{Geophys.\ Res.\ Lett.} 
\newcommand{\iaucirc}{IAU Circ.}         
\newcommand{\iauproc}{Proc.\ of the IAU} 
\newcommand{\ibvs}{IBVS}                 
\newcommand{\icarus}{Icarus}             
\newcommand{\ieeetit}{IEEE Trans.\ Info.\ Theor.} 
\newcommand{\ieeemtt}{IEEE Trans.\ Microwave Theor.\ \& Techniques} 
\newcommand{\ieeemi}{IEEE Trans.\ Med.\ Imaging} 
\newcommand{\ijmpd}{Int'l J.\ Mod.\ Phys.\ D} 
\newcommand{\invp}{Inverse Prob.}        
\newcommand{\jastp}{J.\ Atmos.\ Sol.-Terr.\ Phys.} 
\newcommand{\jcap}{J.\ Cosm.\ Astropart.\ Phys.} 
\newcommand{\jcomph}{J.\ Comput.\ Phys.} 
\newcommand{\jcp}{J.\ Chem.\ Phys.}      
\newcommand{\jewa}{J.\ Electromagn.\ Wav.\ Appl.} 
\newcommand{\jgeod}{J.\ Geodesy}         
\newcommand{\jgr}{J.\ Geophys.\ Res.}    
\newcommand{\jhep}{JHEP}                 
\newcommand{\jrasc}{JRASC}               
\newcommand{\met}{Meteoritics}           
\newcommand{\mmras}{MmRAS}               
\newcommand{\mnras}{MNRAS}               
\newcommand{\moonp}{Moon and Plan.}      
\newcommand{\mpla}{Mod.\ Phys.\ Lett.~A} 
\newcommand{\mps}{Meteoritics and Planetary Science} 
\newcommand{\nar}{New Astron.\ Rev.}     
\newcommand{\nast}{New Astron.}          
\newcommand{\nat}{Nature}                
\newcommand{\nima}{Nucl.\ Instrum.\ Meth.\ A} 
\newcommand{\npbproc}{Nucl.\ Phys.\ B Proc.\ Supp.} 
\newcommand{\njp}{New J.\ Phys.}         
\newcommand{\nspu}{Phys.\ Uspekhi}       
\newcommand{\pasa}{PASA}                 
\newcommand{\pasj}{PASJ}                 
\newcommand{\pasp}{PASP}                 
\newcommand{\phr}{Phys.\ Rev.}           
\newcommand{\pla}{Phys.\ Lett.~A}       
\newcommand{\plb}{Phys.\ Lett.~B}       
\newcommand{\pop}{Phys.\ Plasmas}        
\newcommand{\pra}{Phys.\ Rev.~A}        
\newcommand{\prb}{Phys.\ Rev.~B}        
\newcommand{\prc}{Phys.\ Rev.~C}        
\newcommand{\prd}{Phys.\ Rev.~D}        
\newcommand{\pre}{Phys.\ Rev.~E}        
\newcommand{\prl}{Phys.\ Rev.\ Lett.}    
\newcommand{\pst}{Phys.\ Scr.~T}        
\newcommand{\phrep}{Phys.\ Rep.}         
\newcommand{\phss}{Phys.\ Stat.\ Sol.}   %
\newcommand{\privcom}{priv.\ comm.}      
\newcommand{\procsci}{Proc.\ Sci.}       
\newcommand{\procspie}{Proc.\ SPIE}      
\newcommand{\planss}{Planet.\ Space Sci.} 
\newcommand{\qjras}{QJRAS}               
\newcommand{\radsci}{Radio Sci.}         
\newcommand{\rpph}{Rep.\ Prog.\ Phys.}   
\newcommand{\rqe}{Rad.\ \& Quan.\ Elec.} 
\newcommand{\rgsp}{Rev.\ Geophys.\ Space Phys.\ } 
\newcommand{\rsla}{Phil.\ Trans.\ R.\ Soc.\ A} 
\newcommand{\sal}{Sov.\ Astron.\ Lett.}  
\newcommand{\spjetp}{Sov.\ Phys.\ JETP}  
\newcommand{\spjetpl}{Sov.\ Phys.\ JETP Lett.} 
\newcommand{\spu}{Sov.\ Phys.\ Usp.}  
\newcommand{\sci}{Science}               
\newcommand{\solph}{Sol.\ Phys.}         
\newcommand{\ssr}{Space Sci.\ Rev.}      
\newcommand{\zap}{Z.\ Astrophys.}        

\section{Introduction}

The lunar Askaryan technique, first proposed by Dagkesamanskii \& Zheleznykh~\cite{dagkesamanskii1989}, employs the Moon as a detector of ultra-high-energy (UHE) particles by searching for the Askaryan radio pulse~\cite{askaryan1962} produced when such a particle interacts on the lunar surface.  The large size of the Moon gives such a detector a very large aperture, but the distance to the Moon makes the expected pulse extremely weak, requiring a highly sensitive radio telescope to detect it.  The Square Kilometre Array (SKA)~\cite{carilli2004} will be, when complete, more sensitive in the appropriate frequency range than any current radio telescope by around an order of magnitude~\cite{dewdney2013}, making it highly attractive for this application.  This --- and the precision measurement of air showers with the same instrument~\cite{huege2015} --- is the focus of the SKA High Energy Cosmic Particles (HECP) Focus Group\footnote{\texttt{http://astronomers.skatelescope.org/home/focus-groups/high-energy-cosmic-particles/}}.

Past experiments~\cite{hankins1996,gorham2004a,beresnyak2005,james2010,spencer2010,jaeger2010,buitink2010,bray2014a,bray2011c} have been sensitive only to the UHE neutrino fluxes predicted by certain exotic `top-down' UHE particle production models (e.g.\ \cite{berezinsky2011,lunardini2012}).  While the SKA will not be sufficiently sensitive to detect the lower-energy cosmogenic neutrino flux, it will be able to further constrain top-down models --- and, critically, it will be able to detect the known UHE cosmic ray (CR) flux, providing the first positive detection of a UHE particle with this technique, and allowing it to test the isotropy of the UHECR sky.  For further details of the scientific potential, see the co-contribution by James et al.~\cite{james2015}.

In this contribution, we discuss some of the issues faced by this experiment, the experience gained from past efforts, and the further work to be done before they are overcome.  Section~\ref{sec:ska}, contains a general description of the relevant characteristics of the SKA, while section~\ref{sec:beamforming} contains more details of its beamforming capabilities.  In section~\ref{sec:ionodisp}, we describe the capabilities required to correct for dispersion of an Askaryan pulse as it passes through the ionosphere, and in section~\ref{sec:rfi} we address the issue of spurious signals from radio-frequency interference.  Finally, section~\ref{sec:conc} summarises the outlook for overcoming these issues for successful application of the lunar Askaryan technique with the SKA.

\section{The Square Kilometre Array}
\label{sec:ska}

Phase~1 of the SKA, to be constructed in 2018--2023, will consist of two components.  SKA1-LOW will be an electronically-steered aperture array, similar to the pathfinder instruments LOFAR~\cite{vanhaarlem2013} and MWA~\cite{tingay2012}, with a frequency range of 50--350~MHz and situated in Western Australia.  SKA1-MID will be an array of steerable parabolic dishes, with a selection of receivers to cover the frequency range 350~MHz to 13.8~GHz, located in South Africa.  Phase~2, to be constructed in 2023--2030, is not yet so well defined, but will most likely consist of a large expansion of both of these instruments.

Past lunar Askaryan experiments have used a range of frequencies, from 113~MHz~\cite{buitink2010} to 4.8~GHz~\cite{beresnyak2005}, trading off between the greater Askaryan pulse strength (and hence lower minimum detectable particle energy) at high frequencies and the increased opening angle of the Askaryan emission (and hence greater geometric particle aperture) at low frequencies~\cite{alvarez-muniz2006,scholten2006}.  The HECP Focus Group is focusing exclusively on SKA1-LOW, taking advantage of the latter effect.

SKA1-LOW will consist of 131,072 log-periodic dipole antennas~\cite{diamond2015}, grouped into \mbox{$\sim 450$} stations of 35~m diameter, each with \mbox{$\sim 300$} densely-packed antennas.  The majority of the stations will be grouped within 1~km, forming a dense core to the array, with the remainder spread out over baselines of up to \mbox{$\sim 80$}~km.  The total collecting area will be less than 1~km$^2$; this milestone will be reached with phase~2.

\section{Beamforming}
\label{sec:beamforming}

\begin{figure}
 \centering
 \includegraphics[width=0.45\linewidth]{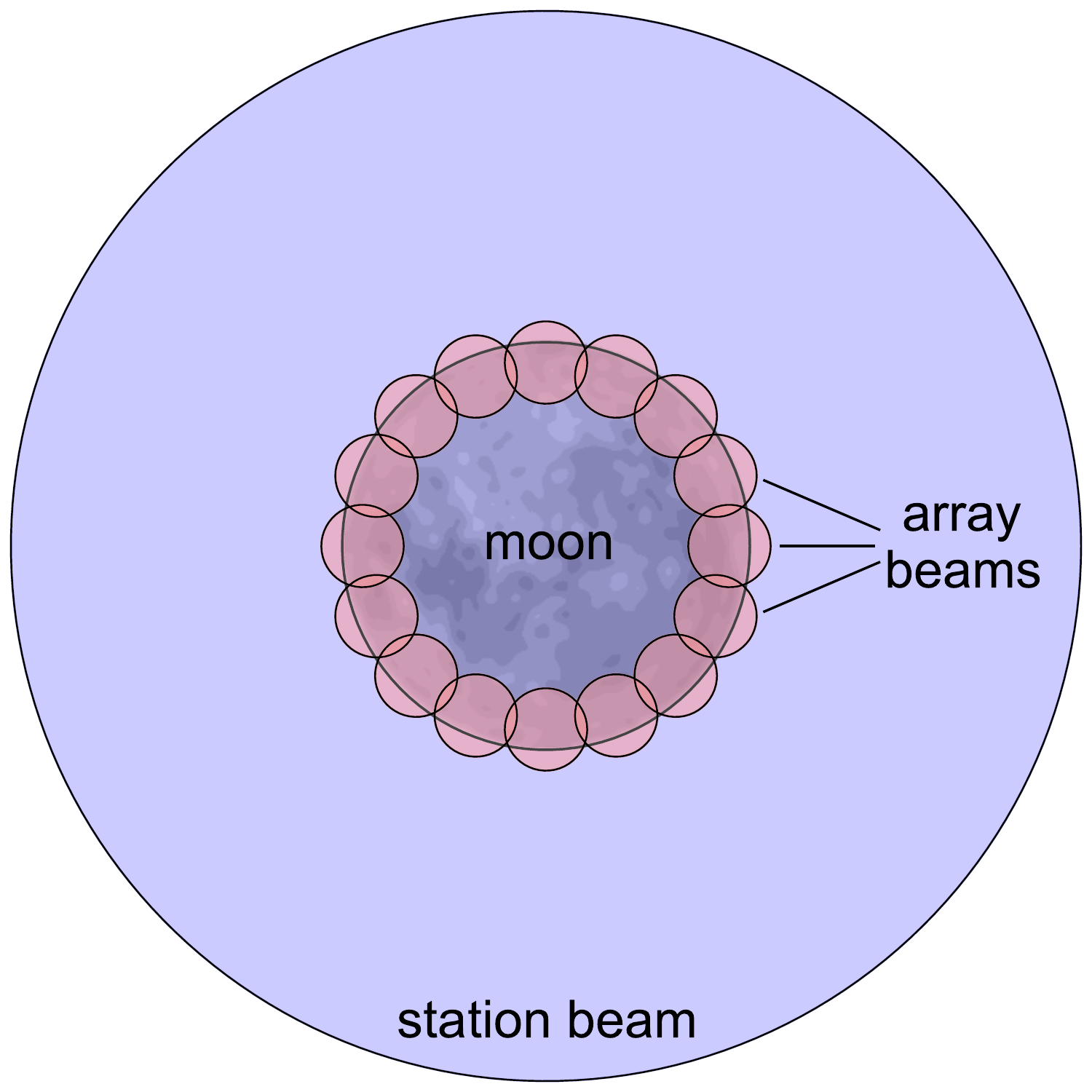}
 \caption{Beamforming strategy for lunar Askaryan observations with the SKA, with the beam from each station centred on the Moon, and all core stations synthesised into multiple array beams directed at different points on the Moon.  These points are selected around the limb of the Moon, from which a signal is most likely to be detected~\cite{james2009f}.  The size of the station beam shown is for the FWHM at the top of the 100--350~MHz band, and the array beams are for the geometric centre of this range.}
 \label{fig:pointing}
\end{figure}

Because the antennas of SKA1-LOW are immobile, pointing of the telescope will be done electronically, with signals from all antennas in a station being summed with appropriate delay offsets to form a station beam pointed at an arbitrary point on the sky.  The full width half maximum (FWHM) size of a station beam has a minimum value of 1.4$^\circ$ at the top of the band, which is sufficient to view the entire Moon, as shown in figure~\ref{fig:pointing}.

\begin{figure}
 \begin{center}
  \begin{minipage}{0.49\linewidth}
   \begin{center}
    \includegraphics[width=0.9\linewidth]{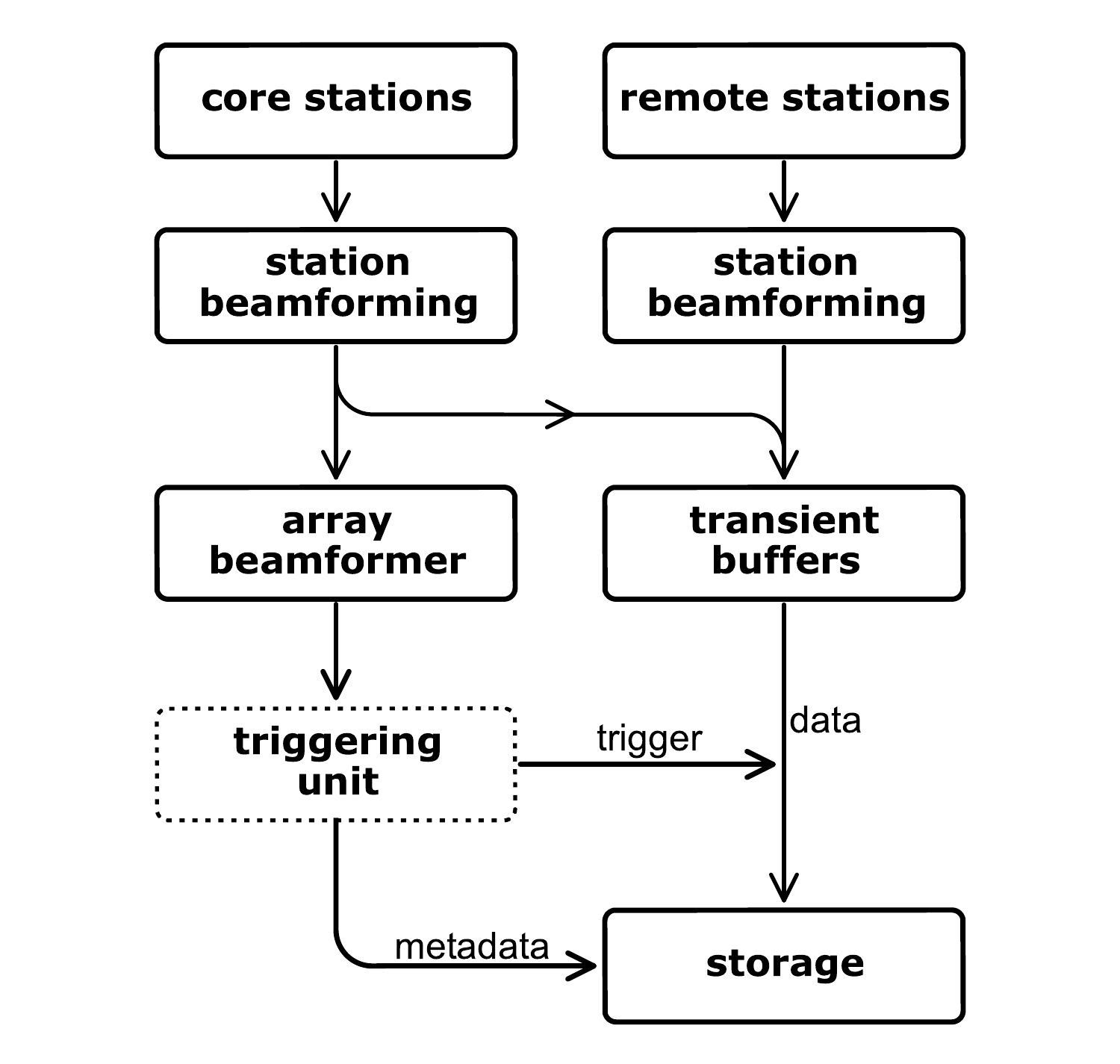} \\[3pt]
    \includegraphics[width=0.9\linewidth]{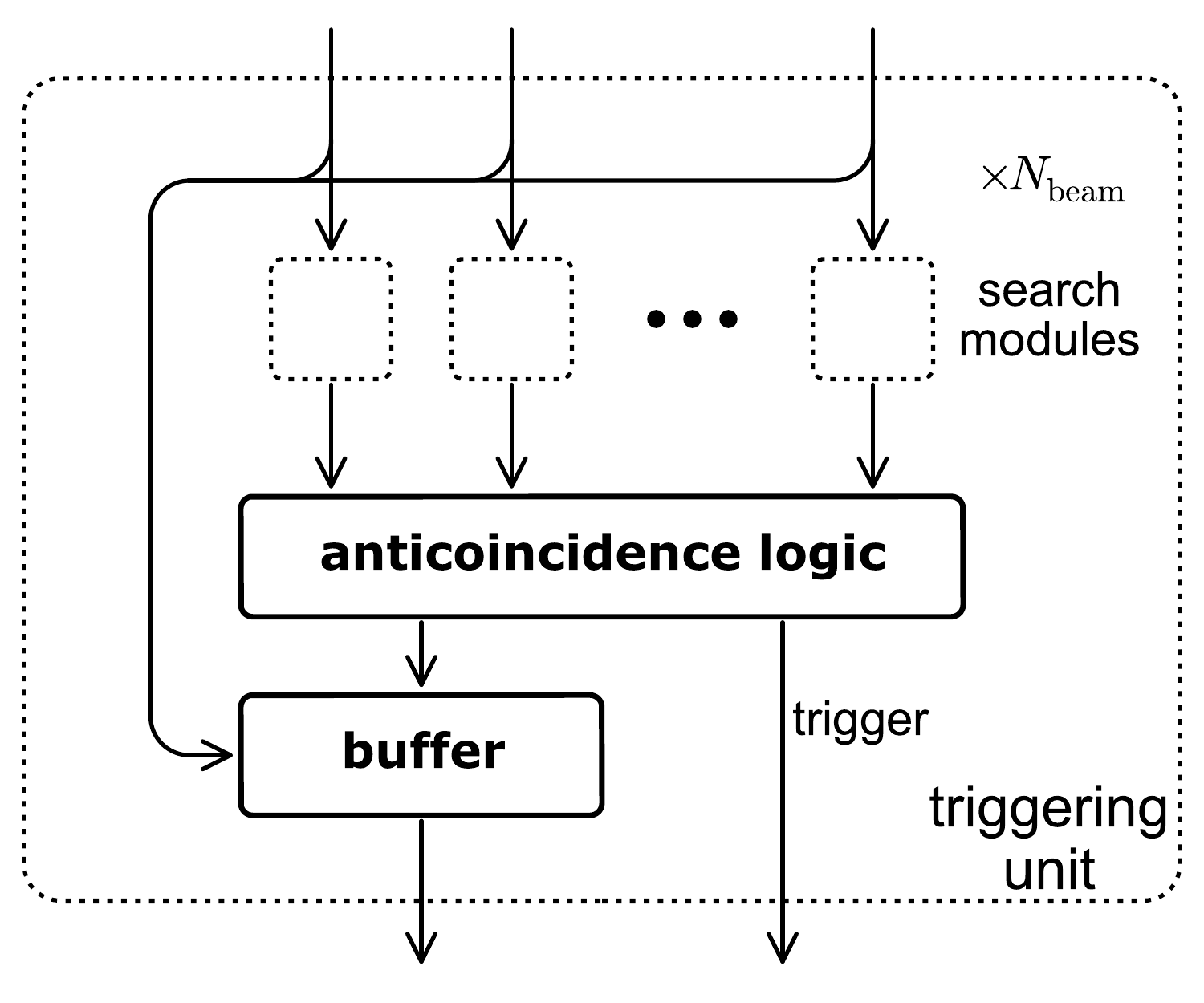}
   \end{center}
  \end{minipage}
  \begin{minipage}{0.49\linewidth}
   \begin{center}
    \includegraphics[width=0.9\linewidth]{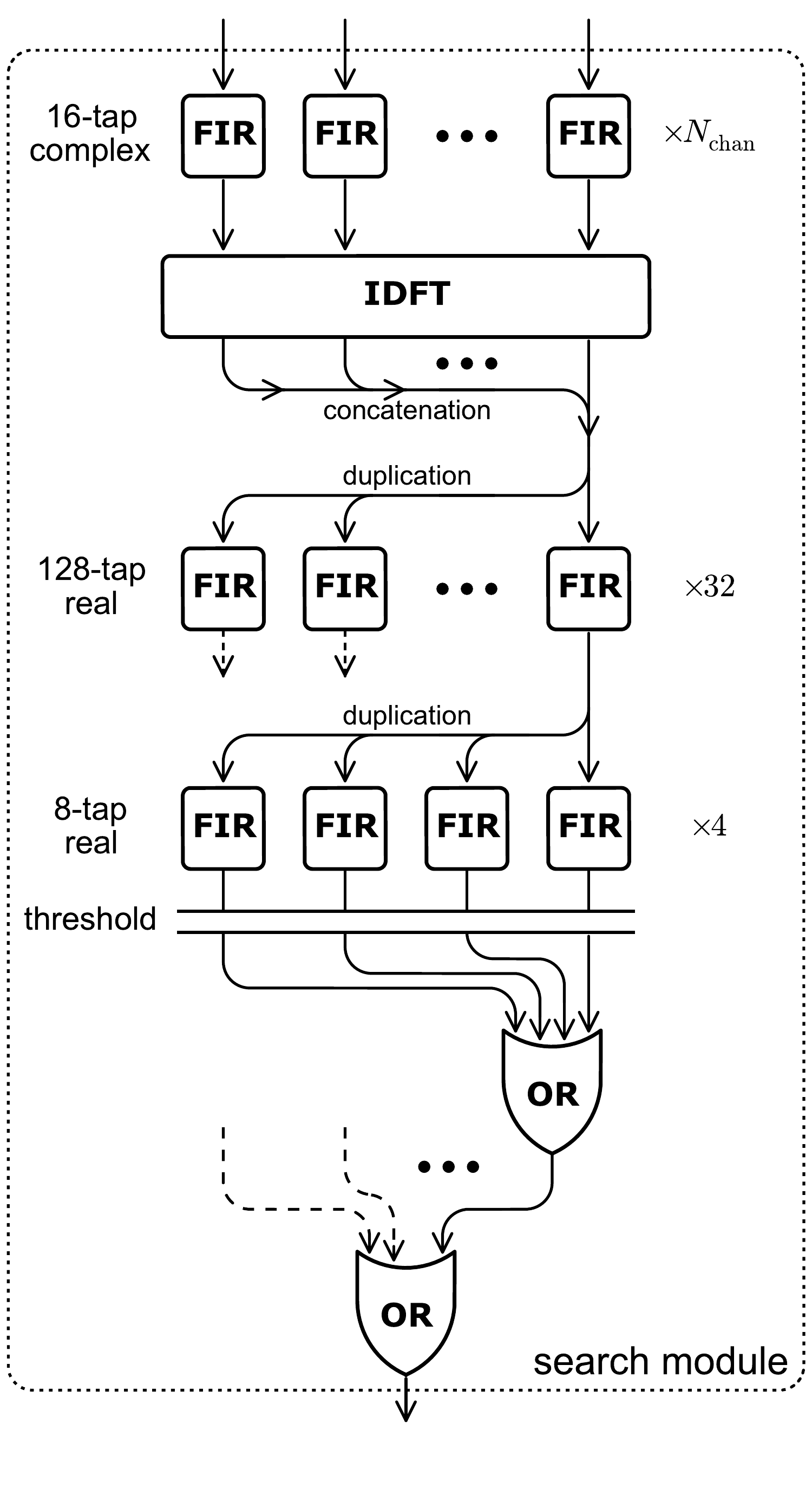}
   \end{center}
  \end{minipage}
 \end{center}
 \caption{Preliminary design of the signal path for a lunar Askaryan experiment with the SKA.  The entire signal path (upper left) includes transient buffers, to be stored when triggered by the detection of a candidate pulse.  The triggering unit (lower left) performs the real-time search for candidates, with each beam being processed by a search module (right).}
 \label{fig:sigpath}
\end{figure}

For high-time-resolution observations such as these, SKA1-LOW will have an array-level beamformer, combining multiple station beams in a similar fashion to form an array beam.  By summing station beams with different delay offsets, this beamformer will be able to form multiple array beams simultaneously, with a planned capacity of 16~dual-polarisation beams with the full bandwidth.  The size and sensitivity of these array beams depends on the set of stations used to form them.  If only stations in the dense core of the array are used, the beams are large enough for them to be tiled completely around the limb of the Moon (see figure~\ref{fig:pointing}), achieving an efficiency factor of \mbox{$\sim 50$}\% of the particle aperture for full lunar coverage, which is incorporated into the simulations of~\cite{james2015}.  The sensitivity is reduced compared to the entire array, but if the array beams are used to trigger storage of buffered data from all stations (per~\cite{bray2014b}; see figure~\ref{fig:sigpath}), the full sensitivity of the array can be achieved retroactively.

The application of a beamformer to the lunar Askaryan technique has already been demonstrated in the NuMoon WSRT experiment~\cite{buitink2010}.  However, the case of SKA1-LOW differs in that the inputs and outputs of this instrument's beamformer are in the form of channelised data produced by a polyphase filter (PPF), so the PPF channelisation must be inverted to restore the original high-time-resolution signal.  This inversion has been shown to be possible and adequately efficient as part of preparatory work for a proposed future NuMoon experiment with LOFAR~\cite{singh2012}.

Retroactively achieving the full array sensitivity requires beamforming in software with stored data.  This capability has been demonstrated, for the detection of slower astronomical transients, by the FRATs project~\cite{terveen2012}.  Of lunar Askaryan experiments, the only one to attempt this thus far is the LUNASKA Parkes-ATCA experiment, which completed some of the required calibration steps~\cite{bray2011c}, but has not demonstrated the capability to form retrospective beams and search them for Askaryan pulses.

\section{Ionospheric dispersion}
\label{sec:ionodisp}

\begin{figure}
 \begin{minipage}{0.48\linewidth}
  \centering
  \includegraphics[scale=0.4]{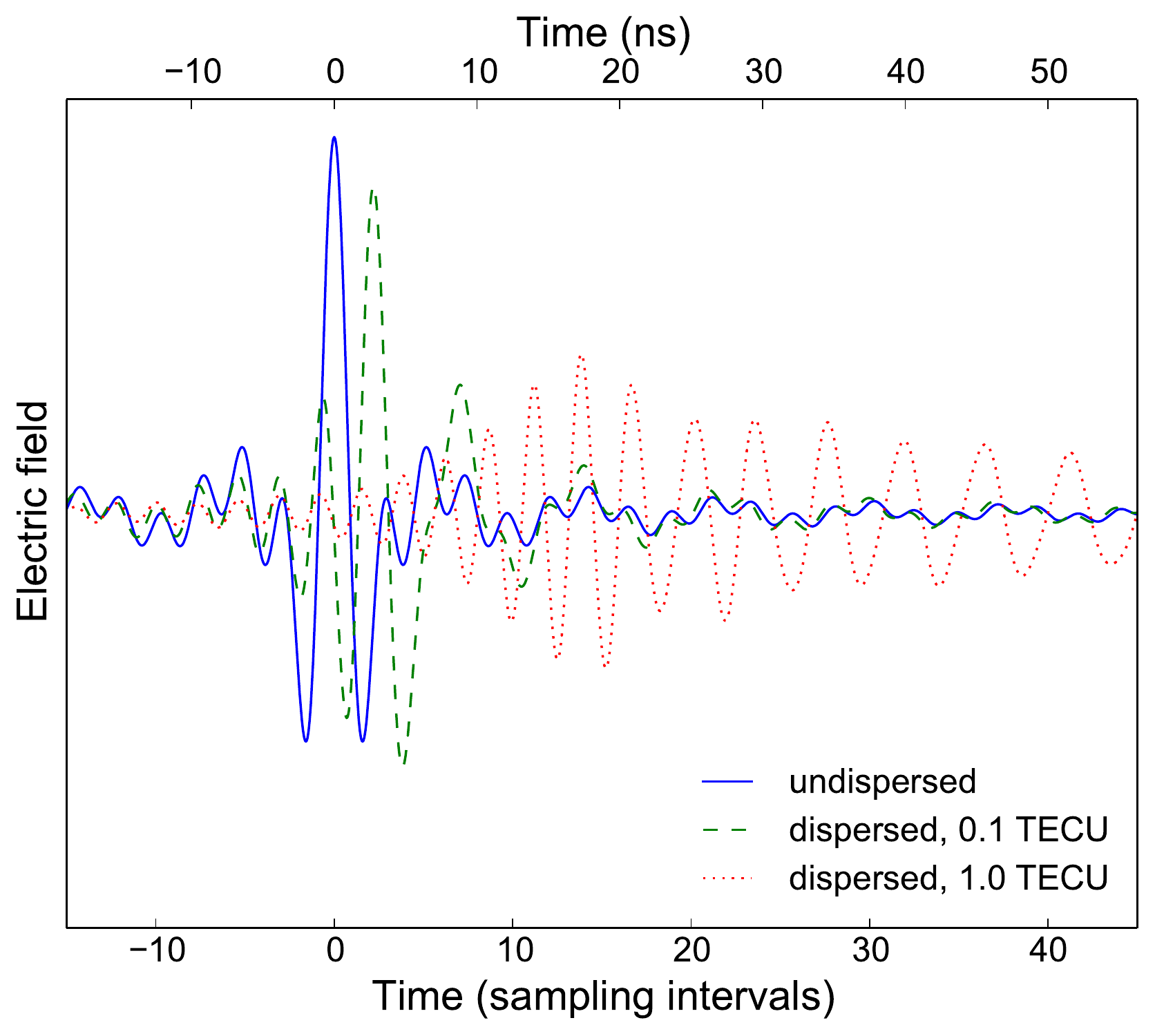}
 \end{minipage}
 \hfill
 \begin{minipage}{0.48\linewidth}
  \vspace{11pt}
  \centering
  \includegraphics[scale=0.4]{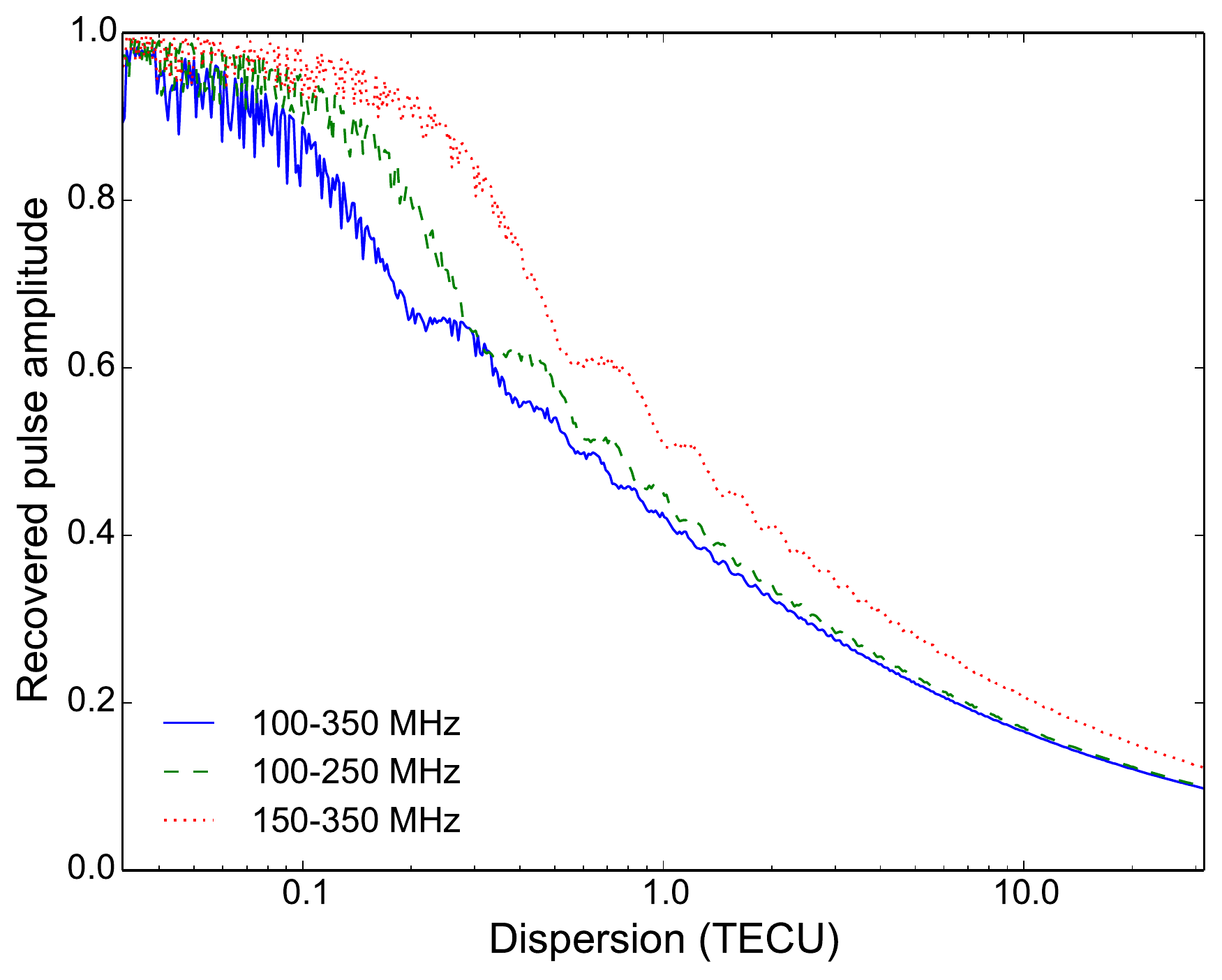}
 \end{minipage}
 \caption{Left: a radio pulse (solid) with a flat spectrum across the frequency range 100--350~MHz, and the same pulse after dispersion equivalent to an electron column density of 0.1~TECU (dashed) and 1.0~TECU (dotted).  Right: the fraction of the original pulse amplitude recovered after this dispersion (or dedispersion error), for several frequency ranges.  A spacing of \mbox{$\Delta{\rm TEC} = 0.1$}~TECU between dedispersion trials ensures that the true dispersion will lie no further than 0.05~TECU from the closest trial, giving a typical recovered pulse amplitude of \mbox{$\gtrsim 95$}\% for any of the frequency ranges shown here.}
 \label{fig:pulsedisp_sensloss}
\end{figure}

A lunar-origin radio pulse propagating through the ionosphere experiences a frequency-dependent delay.  This ionospheric dispersion extends the pulse in time and decreases the peak pulse height, as shown in figure~\ref{fig:pulsedisp_sensloss} (left).  The degree of dispersion is determined by the column density of free electrons, or total electron content (TEC), measured in TEC units (TECU; 1~TECU = $10^{16}$~electrons\,m$^{-2}$).  It also depends on the frequency range of the pulse: the low frequency and broad band of SKA1-LOW make it particularly severe.

Previous experiments have either neglected dispersion, or corrected for the dispersion associated with the measured ionospheric TEC.  For SKA1-LOW, this is insufficient: since the recovered amplitude of the pulse drops so quickly with any error in the TEC (see figure~\ref{fig:pulsedisp_sensloss} (right)), even after excluding the 50--100~MHz band which is most strongly affected by dispersion, a typical uncertainty in the measured TEC of 1--2~TECU~\cite{bray2014a} can result in a loss of over 50\% of the signal amplitude.

\begin{figure}
 \begin{minipage}{0.48\linewidth}
  \centering
  \includegraphics[width=\linewidth]{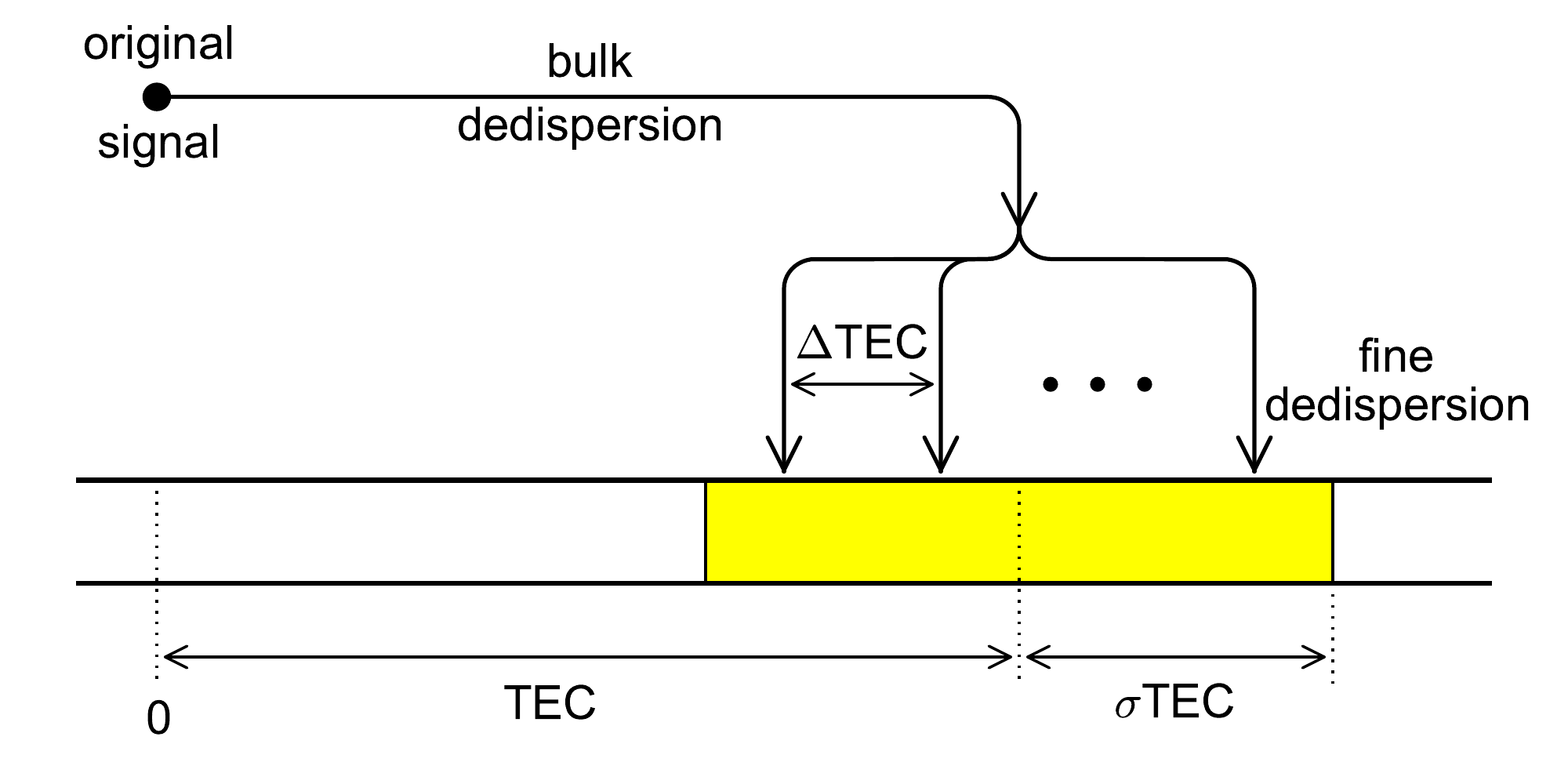}
 \end{minipage}
 \hfill
 \begin{minipage}{0.48\linewidth}
  \vspace{11pt}
  \centering
  \includegraphics[width=\linewidth]{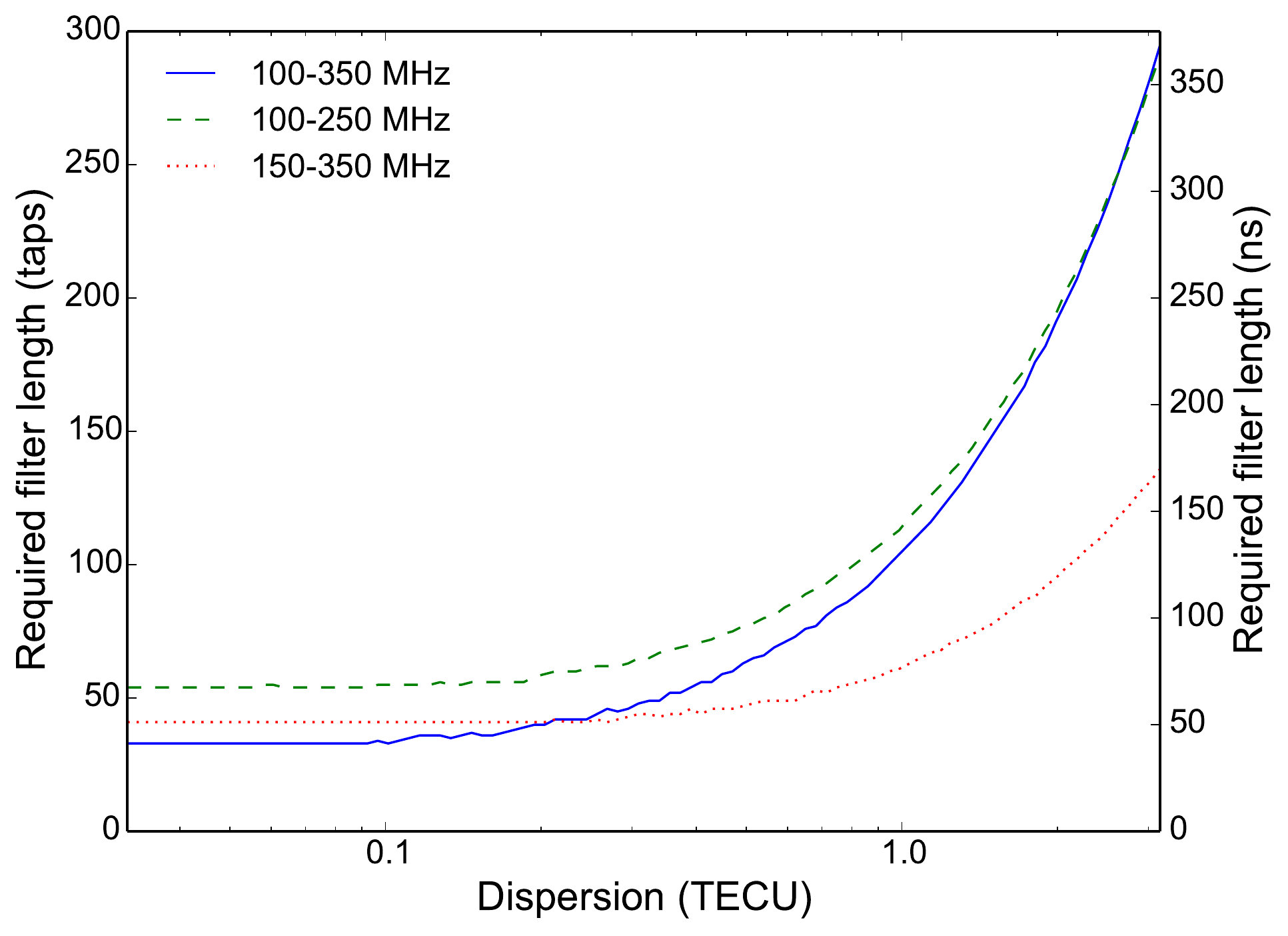}
 \end{minipage}
 \caption{Left: planned dedispersion scheme, with bulk dedispersion compensating for the mean expected TEC, and fine dedispersion to implement a parallel search with spacing $\Delta$TEC across the uncertainty range defined by $\sigma$TEC.  Right: required length of an FIR dedispersion filter to retain \mbox{$>98$}\% of the power of a dispersed template pulse.  To accommodate a dispersion equal to the uncertainty of \mbox{$\sigma{\rm TEC} \sim 1$}~TECU, the required filter length is \mbox{$\sim 100$} taps, depending on the frequency range.}
 \label{fig:dedisp_filterlen}
\end{figure}

A direct solution to this problem is to search in parallel at multiple TECs, as shown in figure~\ref{fig:dedisp_filterlen} (left), at the expense of increased processing requirements.  To span an uncertainty range of \mbox{$\pm 1$}--2~TECU with a spacing between filters of $\Delta{\rm TEC} \sim 0.1$~TECU would require \mbox{$\sim 30$} filters running in parallel.  If implemented as finite impulse response (FIR) filters, each of these would require \mbox{$\sim 100$} taps (operations per sample), per figure~\ref{fig:dedisp_filterlen} (right).  This motivates the 32~parallel FIR filters with 128~taps each in the preliminary search module design shown in figure~\ref{fig:sigpath} (right), which dominate the total computing requirements.

The computational requirements can be considerably reduced if the TEC can be measured more precisely: the number of parallel filters, and the number of operations per filter, are both decreased.  The quoted uncertainty of 1--2~TECU is for two-dimensional TEC maps based on data from global positioning system satellites; direct measurements of the TEC along the line of sight to these satellites have an inherent uncertainty of \mbox{$< 0.1$}~TECU~\cite{hernandez-pajares2009}, and direct interpolation of these measurements should allow something closer to this to be achieved.  Another possibility is to use Faraday rotation of polarised lunar thermal radio emission, which is also determined by the ionospheric TEC, to directly measure the TEC along the line of sight from the telescope to the Moon~\cite{mcfadden2011}.

\section{Radio-frequency interference}
\label{sec:rfi}

A critical problem for a lunar Askaryan experiment is to apply cuts to exclude spurious anthropogenic radio pulses, referred to as radio-frequency interference (RFI), which can appear as a false signal.  In application to the SKA, this problem comes in two parts.  The first is to exclude in real time the majority of RFI, which would otherwise cause an excessively high trigger rate and an unmanageable quantity of stored data.  The second is to completely exclude any remaining RFI in retrospective analysis, so a detected pulse can be identified with high confidence as being a lunar-origin Askaryan pulse produced by a UHE particle.

Past experiments have most commonly used a coincidence requirement falling into one of two categories: requiring a coincident detection by multiple detectors pointed at the same part of the Moon, or requiring an anticoincident detection by only one of multiple detectors pointed at different parts of the Moon.  The former approach is inapplicable here, so we are forced to rely on the second, with the 16~array beams constituting our multiple detectors.  The past experiments that have relied most heavily on this approach (alongside other cuts) are the NuMoon WSRT experiment~\cite{buitink2010} and the LUNASKA Parkes experiment~\cite{bray2014a}.

In the NuMoon WSRT experiment, a substantial number of RFI events passed the cuts, dominating strongly over the thermal noise distribution.  This may be due to the RFI-loud environment of the telescope, the sensitivity of the array beams to the RFI-loud horizon, or the low observing frequency, which typically results in more RFI.  The LUNASKA Parkes experiment had a higher observing frequency and a more RFI-quiet environment, and the cuts applied were sufficient to remove effectively all RFI, but the statistical rigour of this was not clearly established: if a single event had remained after the cuts were applied, it would not be clear if it was a remnant RFI event or a true lunar-origin pulse.

Despite its low frequency, SKA1-LOW has two major advantages over these two experiments.  The first is the telescope site, which is extremely isolated and therefore RFI-quiet.  The second is the availability of station-level buffers which permit retrospective beamforming, through which it is possible to precisely localise a source of RFI in the near field of the telescope.  Together, these advantages make it highly likely that it will be possible to meet the standard of proof required to make a definitive detection of a lunar-origin pulse, but establishing this with certainty will require analysis of stored data from SKA1-LOW in operation.  Until this is possible, a reasonable intermediate step would be the measurement and analysis of the nanosecond-scaled pulsed RFI environment at the telescope site.

\section{Conclusion}
\label{sec:conc}

The Square Kilometre Array provides a potent option for applying the lunar Askaryan technique, with a strong prospect of its first positive detection of an ultra-high-energy cosmic particle, but further technical development is necessary to demonstrate the capability to fully exploit this instrument.  Experiments of this type have not previously searched for Askaryan pulses in retrospectively beamformed data; this capability could be demonstrated through a complete analysis of the LUNASKA Parkes-ATCA experiment, or through the realisation of the planned NuMoon LOFAR experiment.  Current measurements of the ionospheric electron content are sufficient for dedispersion with the SKA, but the computing requirements could be substantially decreased through either direct interpolation of satellite measurements, or measurement of lunar Faraday rotation.  The effectiveness with which false signals from RFI can be excluded cannot be fully established until an experiment is carried out, but extrapolation from previous experiments suggests that it will be practical, and an on-site survey of fast transient RFI would help to demonstrate this.

None of these technical issues appears likely to present an insurmountable obstacle to lunar Askaryan observations with the SKA.  The task now at hand is to demonstrate solutions to them before the scheduled completion of the SKA1-LOW instrument in 2023.

\section*{Acknowledgements}

JDB acknowledges support from ERC-StG 307215 (LODESTONE).  The authors would like to dedicate this contribution to their friend, colleague, and fellow `lunatic' Ray Protheroe, who sadly passed away on Wednesday July 1$^{\rm st}$, 2015.

\newcommand{\titleopt}[1]{}

\newcommand{\arxivopt}[1]{}

\newcommand{\authoropt}[1]{}

\end{document}